\documentclass[11pt,reqno]{article}

\usepackage[parfill]{parskip}  
\usepackage[english]{babel}
\usepackage{amsmath}
\usepackage{amsfonts}
\usepackage{amssymb}
\usepackage{amsthm}
\usepackage[pdftex]{graphicx}
\usepackage[pdftex]{graphics}
\usepackage{epstopdf}
\usepackage{enumitem}
\usepackage{geometry}
\usepackage{hyperref}
\usepackage{tikz}
\usepackage[all]{xy}
\usepackage{multirow}
\usepackage{anysize}
\usepackage{color}
\usepackage{pdflscape}
\usepackage{lineno}

\newcommand{\commentout}[1]{}
\newcommand{\beq}{\begin{equation}}
\newcommand{\eeq}{\end{equation}}
\def\wl{\par \vspace{\baselineskip}}

\marginsize{3.2cm}{3.2cm}{3.2cm}{3.2cm}

\title{The emergence of hyper-altruistic behaviour in conflictual situations}

\author{Valerio Capraro}

\begin{document}


\maketitle
\begin{center}
\emph{Center for Mathematics and Computer Science (CWI)}\\
\emph{1098 XG, Amsterdam}\\ 
\emph{The Netherlands}\\ 
\emph{V.Capraro@cwi.nl}
\end{center}
\wl


\begin{abstract}
Situations where people have to decide between hurting themselves or another person are at the core of many individual and global conflicts. Yet little is known about how people behave when facing these situations in the lab. Here we report a large experiment in which participants could either take $x$ dollars from another anonymous participant or give $y$ dollars to the same participant. Depending on the treatments, participants could also exit the game without making any decision, but paying a cost. Across different protocols and parameter specifications, we provide evidence of three regularities: (i) when exiting is allowed and costless, subjects tend to exit the game; (ii) females are more likely than males to exit the game, but only when the cost is small; (iii) when exiting is not allowed, altruistic actions are more common than predicted by the dominant economic models. In particular, against the predictions of every dominant economic model, about one sixth of the subjects show hyper-altruistic tendencies, that is, they prefer giving $y$ rather than taking $x>y$. In doing so, our findings shed light on human decision-making in conflictual situations and suggest that economic models should be revised in order to take into account hyper-altruistic behaviour.
\end{abstract}

Part of the secret of the enormous success of human societies is our ability to cooperate with others and help less fortunate people\cite{Tr,Ax-Ha,Ra-No,FF03}. Sharing food and cooperating during hunt have played a fundamental role in the early evolution of human societies\cite{KG} and modern variants of these attitudes play a major role still nowadays: we help friends when they need, we make donations to less fortunate people, we collaborate with our partner to build a family, we cooperate with our colleagues to finish the work faster and at higher standards. Lab experiments show that our pro-social abilities go far beyond the five rules of cooperation\cite{No06}: people show pro-social behavior also in one-shot lab experiments with anonymous participants\cite{KKT,Forsythe94,AndreoniMiller,WH,A12,EZ,KL,CJR,CSMN} and even in large groups\cite{BC}.  

A major consequence of our pro-social abilities is that our social network is far more connected than that of any other animal species. While this dense spatial structure has numerous well known positive consequences\cite{Santos,Perc,Rand14e}, it also generates a painful paradox: with all the people we are connected with, it is often difficult to make everyone happy: sometimes the goals of two people are just not aligned; sometimes we have to choose between hurting Person A or hurting Person B; perhaps even worse, sometimes we have to choose between hurting ourselves or hurting someone else - and sometimes, this someone else is a close friend, or a close relative, or our romantic partner. 

Despite the practical importance of such conflicts, little is known about how real people behave in these situations in the ideal scenario of a lab experiment with anonymous subjects. To the best of our knowledge, only one study\cite{Cr14} addressed this problem, showing that most people are ``hyper-altruistic'', that is, they evaluate others' pain more than their own pain: they pay to avoid an anonymous stranger receiving an electric shock twice as much as they pay to avoid themselves receiving an electric shock.

Here we go beyond real physical harm and we show that hyper-altruistic behavior can be observed also in simple economic decisions where no real physical harm is involved. A major upside of this purely economic approach is that it  provides a straight proof that economic models are somehow incomplete, since (as it will be shown in the Results section) they do not predict existence of hyper-altruistic subjects. 

More precisely, here we report experiments on two types of conflicts, those with an exit option and those without an exit option. The typical conflict with no exit option involves two people, person A and person B. Person A has to decide between two allocations of money $(s_1,o_1)$ and $(s_2,o_2)$, the amount $s_i$ being for himself and the amount $o_i$ for Person B. Person B has no active role and only gets what Person A decides to give. The two allocations of money are assumed to be in conflict, that is $s_1>0>s_2$ and $o_1<0<o_2$. Conflicts with an exit option differ from those without an exit option in that Person A can exit the game without making any decisions, but paying an amount $e\geq0$. Thus here Person A has a third choice available, that is the allocation of money $(-e,0)$. We define the \emph{cost} of the exit option to be $c=(e-s_2)-(s_1-e)=2e-s_1-s_2$, that is, the difference between the benefit that Person A would get by exiting the game compared with the worst case scenario, and the loss that Person A would incur if he takes the exit instead of maximising his income. 

We are interested in testing three hypotheses. First, in line with the results presented in ref. 20, we expect to observe hyper-altruistic behavior to a larger extent than predicted by economic models. Second, motivated by the results reported in ref. 21, which show that a substantial proportion of subjects prefer exiting a Dictator game rather than playing it, we expect to see a preference for opting out also in our conflictual situations, at least when the exit option is costless. Third, motivated by the pretty well established fact that females are more giving than males in the Dictator game\cite{EG,AV,DM,HS09,DEJR,D14,KC,B14,BK95,Ca-Ma}, we suspect that there might be gender differences also in behaviour in conflictual situations.

To test these hypotheses, we have conducted three studies (Studies 1, 2, and 3) to explore human behaviour in two-person conflicts with or without an exit option and with different parameter specifications. Details about the designs are reported in the Method section and results are reported in the Results section. Here we anticipate that we have found evidence of four regularities:
\begin{enumerate}
\item[(i)] In the conditions with a costless exit option ($c=0$), the majority of subjects exit the game; 
\item[(ii)] In the conditions with a costly exit option ($c>0$), the majority of subjects act selfishly. Statistically, the proportion of people exiting the game is the same as the proportion of people acting altruistically in the conditions with no exit option.
\item[(iii)] Females are more likely than males to exit the game, but only when the cost of the exit option is small. As the cost of the exit option increases, gender differences in taking the exit option tend to disappear. Moreover, there are no statistically significant gender differences in the conditions with no exit option.
\item[(iv)] In the conditions with no exit option, participants were more altruistic than predicted by the dominant economic models.
\end{enumerate}

Indeed, in the Results section we will show that the observed proportion of \emph{hyper-altruistic} subjects cannot be explained by any of the dominant economic models, including Levine's model of altruism\cite{Levine94}, Fehr \& Schmidt's and Bolton \& Ockenfels' inequity aversion models\cite{Fe-Sc,Bo-Oc}, Charness \& Rabin's efficiency maximisation model\cite{Ch-Ra}, and others\cite{Ha-Pa12,Re-Sc09,HR1,Ca,CVPJ,Ca-Ha}. More precisely, since every participant was asked to describe the reason of his or her choice, with the help of a coder we could analyse the motivation underlying each participant's decision. We have found evidence that hyper-altruistic participants are likely to have some sort of non-consequentialist moral preferences: they either think that taking money from someone else is wrong, or that giving money to someone else is right - independently of the economic consequences.

This finding suggests that increasing the moral weight of the decision problem may have a positive effect on altruistic behavior. In particular, it is possible that taking money from an anonymous person and split it with a third party is perceived to be even ``more wrong'' than just taking money from an anonymous person. Motivated by this observation, we have conducted one more study (Study 4) to investigate whether there is a behavioural transition when passing from two-person conflicts to three-person conflicts.  Here, in the condition with no exit option, Person A has to decide among three allocations of money, $(x,x,-2x), (x,-2x,x)$, and $(-2x,x,x)$, with $x>0$, the first component being for himself and the other two components for Person B and person C, who have no active role. In the condition with an exit option, Person A has a fourth option available, according to which he or she can exit the game at no cost, which corresponds to the allocation $(0,0,0)$. We found the same pattern as in Studies 1, 2, and 3. Most subjects exit the game when the exit option is available and females are more likely than males to exit the game. When no exit option is available, a substantial proportion of subjects act altruistically. However, we found that the frequency of altruistic behavior in this three-person conflict does not significantly differ from the frequency of altruistic behavior observed in the two-person conflicts.

Taken together, our findings shed light on human decision-making in conflictual situations and provide evidence that the dominant economic models should be revised in order to take into account hyper-altruistic behaviour.

\section*{Method}

A total of 2.379 subjects living in the US were recruited using the online labour market Amazon Mechanical Turk (AMT)\cite{PCI,HRZ} and participated in one of four experiments involving money. 

In Study 1, 601 subjects earned $\$0.30$ for participation and were randomly assigned to one of six conditions. In the \emph{no-exit} condition participants were asked to decide between \emph{stealing} Person B's participation fee or \emph{donating} their participation fee to Person B. Subjects in the role of Person B participated in the \emph{guess-no-exit} condition and they had to guess Person A's decision with a $\$0.10$ reward in case they made the right guess. The \emph{free-exit} and \emph{guess-free-exit} conditions were similar, with the difference that there was a third choice available to Person A, that is, exit the game without doing anything. In this case both subjects would keep their participation fee. Finally, the \emph{costly-exit} and \emph{guess-costly-exit} conditions differed from the free-exit conditions in that exiting the game costed $\$0.05$ to Person A. After making their decision, participants entered the demographic questionnaire, where we asked for their gender, age, and education level, and the reason of their choice. Full instructions are reported in the Supplementary Information.

Since AMT does not allow experimenters to manipulate participation fees, Study 1 actually involves deception: participants' choices did not have a real impact on their final bonus. Moreover, one may contest the use of the verb ``to steal'', which, having a strong moral weight, might have driven some participants away from selfish behaviour for other reasons than their altruism. Analysing participants' free responses to the question ``Why did you make your choice?'', we did not find any evidence that participants were aware of the risk of deception; however, we have found evidence that the use of the verb ``to steal'' may have affected participants' choices. Indeed, several participants, when describing their choice, declared ``I am not a thief'', or similar statements. 

To exclude the risk that our results were driven by either of those two causes, Study 2 replicates the no-exit condition of Study 1 under slightly different conditions. 

Specifically, in Study 2, 583 subjects kept their participation fee and were given additional $\$0.30$ as a bonus. Then participants in the role of Person A were asked to decide between \emph{taking} the other participant's bonus or \emph{giving} their bonus to the other participant. Full instructions are reported in the Supplementary Information.

Observing altruistic behaviour in the no-exit condition of Study 1 and in Study 2 will allow us to conclude that there are some subjects who care about the payoff of the other person \emph{at least as much} as their own. The purpose of Study 3 (395 subjects) is to strengthen this conclusion showing that a substantial proportion of subjects is \emph{hyper}-altruist: they care about the payoff of the other person \emph{more} than their own. Thus in Study 3, participants kept their participation fee, were given additional $\$0.10$, and were randomly assigned to either the exit-condition or the no-exit condition. In the no-exit condition, participants in the role of Person A were asked to decide between giving their money to the other person or taking the money from the other person. In the latter case, the money would be doubled and earned by themselves. The exit condition was very similar, a part from the fact that participant were allowed to exit the game without making any decision and paying any cost. Full instructions are reported in the Supplementary Information.

Finally, Study 4 (600 subjects) investigates a three-person conflict with or without costless exit option. Here, participants kept their participation fee, were given additional $\$0.30$, and were randomly assigned to either the exit-condition or the no-exit condition. In the no-exit condition, participants in the role of Person A were asked to decide between giving their money to two other people ($\$0.15$ each) or taking one of these people's $\$0.30$ and splitting it with the third person. The exit condition was very similar, a part from the fact that participants were allowed to exit the game without making any decision and paying any cost. Full instructions are reported in the Supplementary Information.

After collecting the decisions, bonuses were computed and paid. These experiments have been conducted in July 2014, while the author was still employed by the University of Southampton, United Kingdom. Informed consent was obtained by all participants. These experiments were approved by the Southampton University Ethics Committee on the Use of Human Subjects in Research and carried out in accordance with the approved guidelines. 

\section*{Results} 

\emph{Study 1.} 

We start by analysing the choices made by the participants who played in the role of Person A. Figure 1 reports the relevant results. In the no-exit condition, $28\%$ of the 101 subjects decided to donate their participation fee. Adding the possibility to exit the game for free had the effect that most participants took the exit. Specifically, $70\%$ of the 100 subjects who participated in the free-exit condition decided to exit the game, while all but three of the remaining participants acted selfishly. Three people preferred to donate their participation fee. The fact that virtually nobody acted altruistically in the free-exit condition also shows that the results of the no-exit condition were not driven by people who did not understand the rules of the game. The costly-exit condition gave statistically the same results as the no-exit condition: $30\%$ of the participants chose to exit the game; all but four of the remaining ones acted selfishly; four people donated their participation fee. In all three conditions, we found that females were more likely than males to act altruistically, although the effect was nearly significant only in the two conditions with an exit option (Rank-sum, $p=0.5353, p=0.0488, p=0.0615$, respectively). The fact that this effect is only marginally significant is due to the relatively small sample size: aggregating over the exit conditions we found that females were highly more likely than males to exit the game ($67\%$ vs $42\%$, $p=0.0048)$.

\begin{figure} 
   \centering
   \includegraphics[scale=0.72]{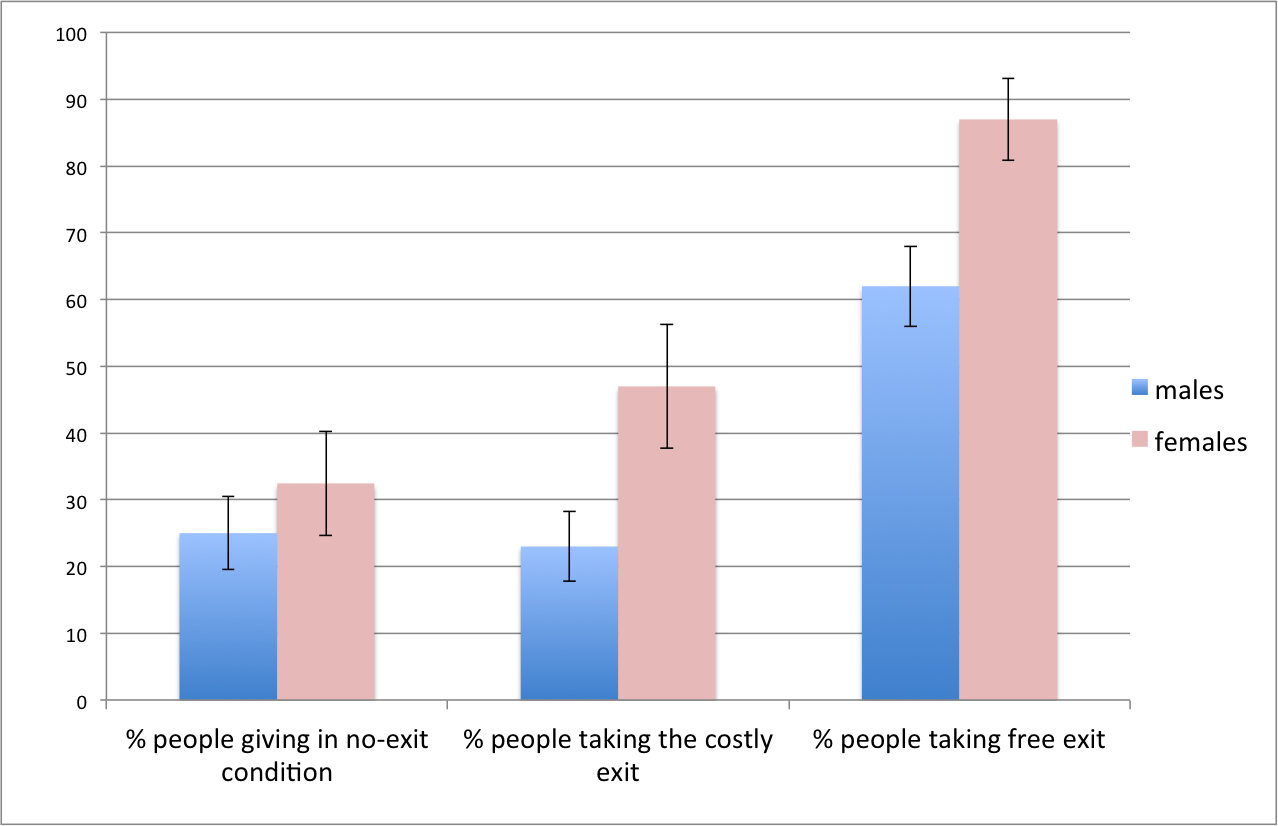} 
   \caption{\emph{In the no-exit condition, about 28\% of subjects preferred giving $\$0.30$ to an anonymous person, rather than taking the same amount of money from that person. Error bars represent the standard error of the mean. Females tended to give more, though the difference was not statistically significant. In the costly-exit condition, about 30\% of subjects preferred paying $\$0.05$ to exit the game without making any decision, rather than making a decision. Females were more likely than males to exit the game ($p=0.0615$).  In the free-exit condition, most subjects preferred to exit the game without making any decision and without paying any cost. Females were more likely than males to exit the game ($p=0.0488$). The p-values are only nearly significant, but this is also due to the small sample size. Aggregating over both exit conditions, we find $p=0.0048$.}}
   \label{fig:conflict}
\end{figure}

Looking at the choices made by the participants who played in the role of Person B, we found that people made statistically the right guess in the guess-no-exit and the guess-costly-exit conditions, while they significantly underestimated the percentage of people taking the exit in the guess-free-exit condition. Specifically, $24\%$ of the 102 subjects in the guess-no-exit condition bet on Person A's giving, compared with the $28\%$ of subjects that actually chose to give in the exit condition ($p=0.6241)$; $30\%$ of the 99 subjects in the guess-costly-exit bet on Person A taking the exit, compared with the same percentage that actually took it in the exit condition ($p=0.9281$); and $49\%$ of the 99 subjects in the guess-free-exit condition bet on Person A taking the exit, compared with the $70\%$ of subjects that actually took it ($p=0.0083$).
\wl
\emph{Study 2.}

Study 2 is a replication of the no-exit condition of Study 1 with slightly different experimental instructions. A total of 583 subjects participated in Study 2 in the role of Person A. The results show no significant difference with the no-exit condition in Study 1: some $21\%$ of the participants preferred giving their money away rather than taking it from the other participant. This percentage does not significantly differ from that in the no-exit condition in Study 1 (Rank sum, $p=0.2891$). Again, females were slightly more altruistic than males ($27\%$ vs $18\%$, $p=0.0873$). This suggests that participants in Study 1 were not aware of the risk of deception and that the use of the non-neutral verb ``to steal'' had a very little effect on participant's choices, if any.
\wl
\emph{Study 3.}

A total of 395 subjects participate in our Study 3 in the role of Person A. Figure 2 reports the relevant results. In the no-exit condition, $17\%$ of the 198 subjects, preferred the allocation $(\$0,\$0.20)$ over $(\$0.30,\$0)$. In the exit-condition, 13 subjects chose to act altruistically, despite the presence of the exit. Among the remaining 184 subjects, only $28\%$ of the subjects took the exit option. There is clearly no gender differences in either conditions. Observe that the cost of the exit option is $\$0.10$ in Study 3, compared with $c=\$0.05$ in the costly-exit condition of Study 1 and $c=0$ in the free-exit condition of Study 1 and in the exit condition of Study 4. Thus this provides evidence that, as the cost of the exit option increases, fewer and fewer people take the exit option and gender differences in taking the exit option tend to disappear. 

\begin{figure} 
   \centering
   \includegraphics[scale=0.65]{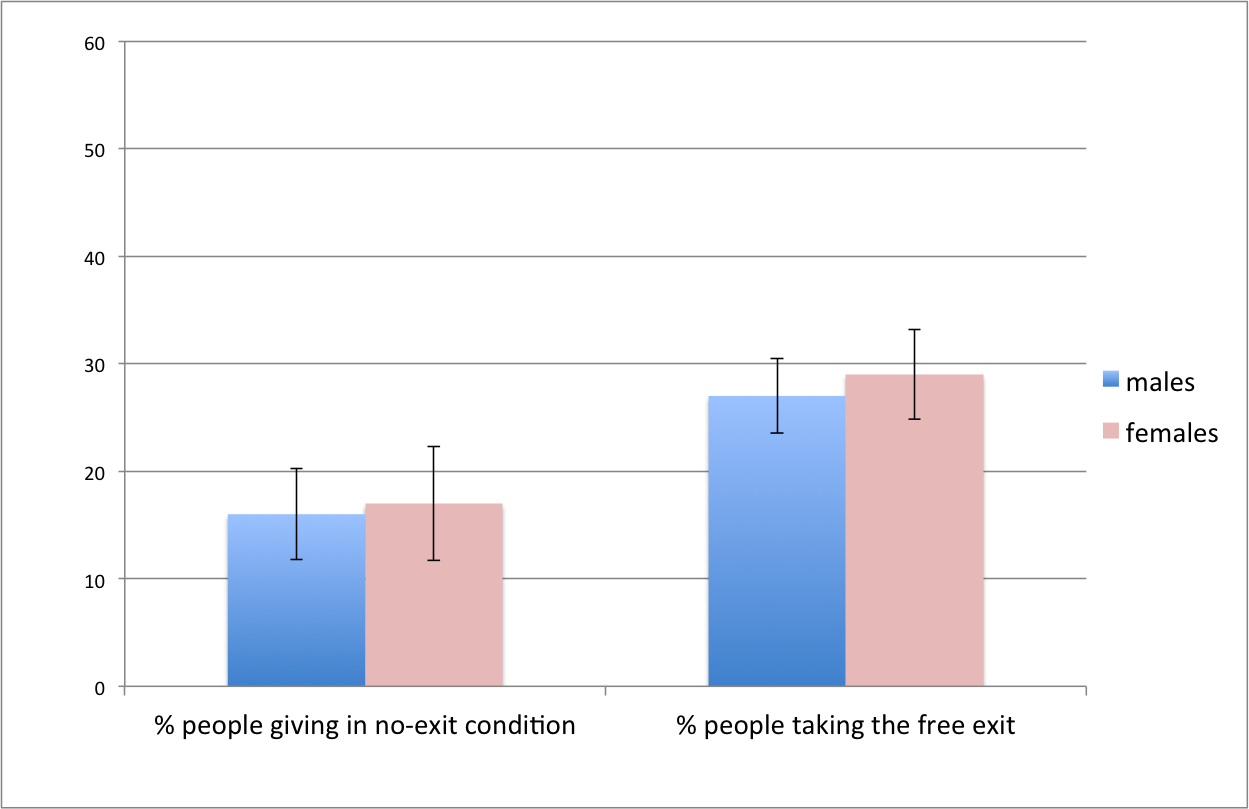} 
   \caption{\emph{In the no-exit condition, about 17\% of subjects preferred the allocation $(\$0,\$0.20)$ over $(\$0.30,\$0)$. Error bars represent the standard error of the mean. In the exit condition, 13 subjects acted altruistically and are not reported in the figure. Among the remaining participants, only $28\%$ of them took the exit. There is clearly no gender differences in either conditions.}}
   \label{fig:no-charness}
\end{figure}
\wl
\emph{Study 4.}

A total of 600 subjects participated in our Study 4, where participants were asked to make a decision in a three-person conflict instead of a two-person conflict as in Studies 1, 2, and 3. Figure 3 reports the relevant results. Perhaps contrary to the expectations, we did not find any significant difference between three-person conflicts and two-person conflicts. In the no-exit condition, $28\%$ of the subjects opted for the altruistic action, while the remaining ones chose either of the selfish options at random. Again, we found that females were slightly more altruist than males ($33\%$ vs $24\%$), though, again, the difference is not statistically significant ($p = 0.1675$). Among the 299 subjects who participated in the free-exit condition, 21 (11 males) chose the altruistic choice, regardless the existence of the way out. Among the remaining 278 subjects, $59\%$ chose the way out. Again we found that females were significantly more likely than males to exit the game ($69\%$ vs $52\%$, $p = 0.0131$).

\begin{figure} 
   \centering
   \includegraphics[scale=0.65]{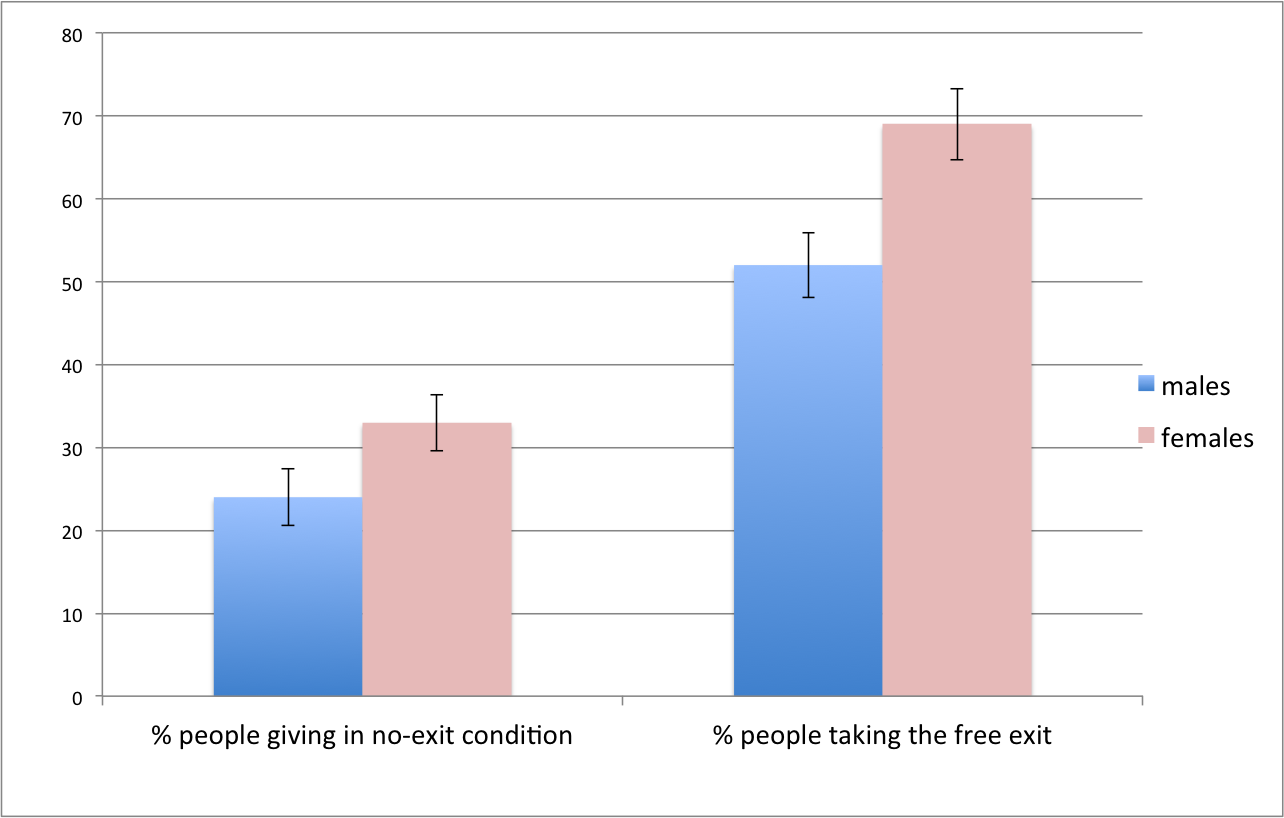} 
   \caption{\emph{In the three-person no-exit condition, about 28\% of subjects preferred giving $\$0.30$ to two anonymous people ($\$0.15$ each), rather than taking the same amount of money from one of these people and sharing it with the third one. Error bars represent the standard error of the mean. Females tended to give more, though the difference was not statistically significant. In the free-exit condition, about 59\% of subjects preferred to exit the game without making any decision and without paying any cost. Females were significantly more likely than males to exit the game ($p=0.0131$).}}
   \label{fig:three-conflict}
\end{figure}
\wl

\emph{Distribution of choices in the conditions with an exit option}

Figure 4 summarizes the distribution of choices in the conditions with an exit option. Subjects tend to exit the game only when the exit option is costless. Even for exit options with a small cost ($c=\$0.05$ in Study 1 and $c=\$0.10$ in Study 3), behaviour seem to reverse: the majority of people act selfishly. Across all conditions, we note a small percentage of people, ranging from 3\% to 7\%, who acted altruistically, despite the presence of an exit option. The nature of these people is at the moment unknown. The analysis of participants' free responses (we asked the participants to describe their choice in Study 1 and Study 3, but not in Study 2 and Study 4) suggests that some of these people did not understand the rules of the decision problem. Interestingly, the remaining ones described themselves as particularly generous. However, the total number of people making this choice is so small that at the moment it is impossible to draw general conclusions.  

\begin{figure} 
   \centering
   \includegraphics[scale=0.65]{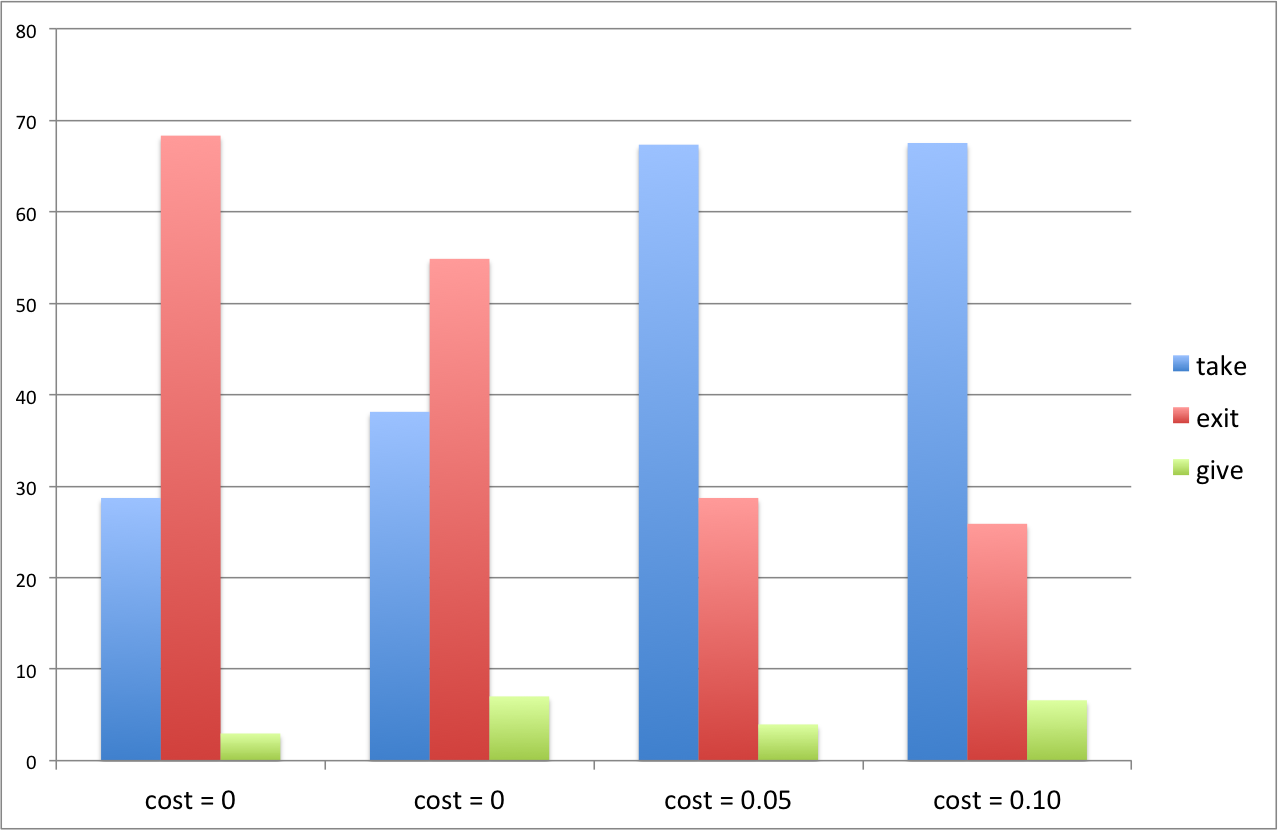} 
   \caption{\emph{Distribution of choices in the condition with an exit option. When the exit option is costless, the majority of people take the exit. This positive effect of the exit option vanishes as soon as participants are asked to pay to exit the game. In this case, the majority of people remain in the game and act so as to maximise their payoff. In all conditions, a small percentage of people, ranging from 3\% to 7\% acted altruistically, despite the presence of the exit option.}}
   \label{fig:distribution}
\end{figure}
\wl

\emph{Economic models do not predict hyper-altruistic behavior}

Following Crockett et al., we say that a person is \emph{hyper-altruist} if he evaluates others' payoff more than his own\cite{Cr14}. Formally, this corresponds to saying that a person \emph{strictly} prefers the allocation of money $(0,y)$ over $(x,0)$, for some $x\geq y$, where the first component is for himself and the second component for an anonymous stranger he is matched with. In this section we show that
\begin{enumerate}
\item About one-sixth of our subjects acted hyper-altruistically;
\item None of the dominant economic models predict existence of hyper-altruistic people.
\end{enumerate}

We note that the first statement is not an obvious consequence of our experimental results, since it might be possible that some subjects are indifferent between $(x,0)$ and $(0,y)$. Half of these subjects would statistically choose the allocation $(0,y)$. As it will be shown later, this behavior would be consistent with Bolton \& Ockenfels' inequity aversion model\cite{Bo-Oc} and with Charness \& Rabin's efficiency maximisation model\cite{Ch-Ra}. However, we now show that this is not case: virtually all people who chose $(0,y)$, made this choice because they strictly preferred $(0,y)$ over $(x,0)$; this means that about one sixth of the total of our subjects acted hyper-altruistically.

To do so, we asked a  research assistant to code each response from the altruistic participants in Study 3. The coder was not informed about the purpose of the study and the hypothesis and predictions being tested. For each statement, she was asked which of the following five categories best described it:

\begin{enumerate}
\item[] The participant explicitly said that they took the action because that was the right thing to do.
\item[] The participant explicitly said that they took the action because the other action was wrong.
\item[] The participant explicitly said that they took the action because they are generous.
\item[] The participant explicitly said that they took an action at random, because they were indifferent between the two actions.
\item[] The participant said something that is not classifiable in any of the previous categories.
\end{enumerate}

We excluded from these analysis three participants who left the free response blank. Among the remaining 30 hyper-altruistic subjects, our coder reported that only two responses can be classified as ``indifferent'' and two responses are not classifiable, while all remaining responses belong to one of the first three categories. More precisely, 8 responses were classified in the ``rightness'' category, 10 responses were classified in the ``wrongness'' category, and 8 responses were classified in the ``generosity'' category. Full classification is reported in the Supplementary Information.

These results unambiguously show that hyper-altruistic participant were not indifferent between the two choices. They acted in a hyper-altruistic way, because ``giving is right'', or ``taking is wrong'', or because they felt generous. 

After showing that hyper-altruism exists and have driven our results, we show that four of the best known economic models of human behaviour do not predict existence of hyper-altruistic behavior. Additionally, we show that also the other (less known) models we are aware of do not predict existence of hyper-altruistic behavior. While we cannot exclude the existence of a model predicting our results, we are now aware of any of them. 

Consider the following decision problem. Let $0<y\leq x$ be fixed, Person A has to decide between the allocation of money $(x,0)$ and $(0,y)$,the first component being for himself and the second component for Person B. Person B has no active role and only gets what Person A decides to give.

We start by analysing the predictions of Levine's model of altruism\cite{Levine94}. This model assumes that, given an allocation of money $(x_1,x_2)$, Player 1 gets an utility of

$$
u_1(x_1,x_2)=x_1+\frac{a_1+\lambda a_2}{1+\lambda}x_2,
$$
where $0\leq\lambda\leq1$ and $-1<a_1,a_2<1$. In particular, the second condition means that no player has a higher regard for his opponents than for himself. It is easy to see that this property implies that Person A strictly prefers the allocation $(x,0)$ over $(0,y)$, independently of the parameters of the model. Indeed $u_1(x,0)=x\geq y>(a_1+\lambda a_2)y/(1+\lambda)=u_1(0,y)$. This prediction is rejected by the results of Studies 1,2, and 3.

We now consider Fehr \& Schmidt's inequity aversion model\cite{Fe-Sc}. This model assumes that, given an allocation of money $(x_1,x_2)$, Player 1's utility is

$$
u_1(x_1,x_2)=x_1-\alpha_1\max(x_2-x_1,0)-\beta_1\max(x_1-x_2,0),
$$

where $\beta_1\leq\alpha_1$ and $0\leq\beta_1<1$. In our case, we have $
u_1(x,0)=x-\beta_1x$ and $u_1(0,y)=-\alpha_1y$. Now assume $x=y$, as it is in Studies 1 and 2, since $\beta_1\leq\alpha_1$, it follows that every decision maker prefers $(x,0)$ over $(0,x)$. This prediction is rejected by the results of Studies 1 and 2. 

Then we consider Bolton \& Ockenfels' inequity aversion model. This model assumes that, given an allocation of money $(x_1,x_2)$, with $x_1+x_2>0$, Player 1's utility is 

$$
u_1(x_1,x_2)=\alpha_1x_1-\frac{\beta_1}{2}\left(\frac{x_1}{x_1+x_2}-\frac{1}{2}\right)^2,
$$

where $\alpha_1\geq0$ and $\beta_1>0$ are constant. Thus, in our case, we have $u_1(x,0)=\alpha_1x -\beta_1/8$ and $u_1(0,y)=-\beta_1/8$, which implies $u_1(x,0)\geq u_1(0,y)$. Consequently, Bolton \& Ockenfels' model predicts that every player either prefers the allocation $(x,0)$ or she is indifferent between the two allocations $(x,0)$ and $(0,y)$; in other words, no player strictly prefers $(0,y)$ over $(x,0)$. This prediction is rejected by the analysis of participants' free responses, which show that essentially all altruistic subjects were actually hyper-altruistic, that is, they strictly prefer $(0,y)$ over $(x,0)$.

Finally, we consider Charness \& Rabin's model\cite{Ch-Ra}, which assumes that, given an allocation of money $(x_1,x_2)$, Player 1's utility is

$$
u_1(x_1,x_2)=(1-\alpha_1)x_1+\alpha_1(\beta_1\min(x_1,x_2)+(1-\beta_1)(x_1+x_2)),
$$

where $\alpha_1,\beta_1\in[0,1]$. Thus we have $u_1(x,0)=(1-\alpha_1)x+\alpha_1(1-\beta_1)x$ and $u_1(0,y)=\alpha_1(1-\beta_1)y$. Since $x\geq y$, one always have $u_1(x,0)\geq u_1(0,y)$. Thus also Charness \& Rabin model predicts that no players strictly prefer $(0,y)$ over $(x,0)$. The argument discussed above applies also in this case and rejects this prediction.

We conclude by mentioning that also the other models of human behaviour in one-shot simultaneous-move games that we are aware of fail to predict existence of hyper-altruistic behaviour. Halpern \& Rong model\cite{HR1} assumes that people care also about the total welfare. However, this model reduces to the money maximisation model in case the total welfare is constant across choices and thus it predicts that players should always choose the allocation $(x,0)$ over $(0,x)$. Similarly, also the cooperative equilibrium model\cite{Ca,CVPJ,BC} reduces to the money maximisation model in case the total welfare is constant across choices. Regret minimisation models\cite{Ha-Pa12,Re-Sc09} instead assume that players compare the payoff obtained when a certain strategy profile is played with the best payoff they could have gotten choosing another strategy and leaving the strategies of the other players constant. Then they try to minimise this \emph{regret}. It is evident that also this model predicts that every player should prefer the allocation $(x,0)$ over $(0,x)$. Finally, the recently proposed model with translucent players\cite{Ca-Ha}, which is based on the illusion of transparency\cite{Gilovich}, that is the illusion that people's thoughts are visible to other people (who can respond punishing unfair intentions), also reduces to the money maximization model in the case in which the other players have no active role and so they cannot punish.

\section*{Discussion}

Here we have provided evidence of the following three regularities: (i) a substantial proportion (about one sixth) of people is hyper-altruist, that is, they prefer giving a certain amount of money to an anonymous stranger, rather than taking the same amount of money from the same person; (ii) the majority of people prefer to avoid this conflictual decision and exit the game, but only when the exit-option is costless; (iii) females are more likely than males to exit the game, even when it is costly, but this gender difference tend to vanish when the cost of the exit option increases.

Existence of hyper-altruism is certainly our major result, since it is not predicted by any of the dominant economic models, including Levine's model of altruism\cite{Levine94}, Fehr \& Schimdt's and Bolton \& Ockenfels' inequity aversion models\cite{Fe-Sc,Bo-Oc}, Charness \& Rabin's efficiency maximisation model\cite{Ch-Ra}, and others\cite{Ha-Pa12,Re-Sc09,HR1,Ca,CVPJ,Ca-Ha}. We are not aware of any model predicting existence of hyper-altruistic people. As a consequence, it is important to understand what psychological and economic motivations led a substantial percentage of people away from the theoretical predictions. Our results provide a starting point in that they suggest that hyper-altruistic behaviour is driven by three different (though probably connected) forces: desire to do the right thing; desire not to do the wrong thing; desire to be generous. Further research is necessary to understand how these forces interact and how they can be incorporated into a model of human behaviour.

A recent paper\cite{Cr14} makes a point similar to our point (i). There, Crockett et al. show that most people evaluate others' pain more than their own pain: they pay to avoid an anonymous stranger receiving an electric shock twice as much as they pay to avoid themselves receiving an electric shock. Though similar, our results are different in the way that they point out that there is no need of real physical harm to observe hyper-altruistic behaviour. In our experiment, a substantial proportion of people value others' monetary outcome more than their own, without any real physical harm involved.

Another paper\cite{Dana} makes a point similar to our point (ii), that is that most people prefer to exit the game, rather than making a decision that would harm either of the parties. There the authors show that about 28\% of subjects prefer to exit a dictator game with \$9, rather than playing it in the role of the dictator with an endowment of \$10. More precisely, participants in ref. 21 played a two-stage game: Stage 1 was
a standard Dictator game where participants in the role of the dictator had to decide how
to allocate \$10 between them and an anonymous recipient, knowing that the recipient
would not have any active role. After making the decision, but before telling it to the
recipient and before telling to the recipient that they were playing a Dictator game in
the role of the recipient, the dictators played Stage 2, in which they were asked whether they wanted to stick with
their decision or leave the game with \$9. In this latter case, the recipient would not
be informed of the fact that they were supposed to be the recipient in a Dictator game.
The authors found that 11 subjects (corresponding to 28\% of the total) preferred to exit the game. Our results extend this finding to conflictual situations and they also make a little step forward: in ref. 21, only two of the 11 subjects who decided to exit the game had decided to keep the whole endowment for themselves in the first
stage of the game. Thus it is possible that the fact that the strategy space has been
changed from Stage 1 to Stage 2, and the fact that the recipient does not have complete
knowledge of the decision problem, have changed some people's preferences, which in Stage 2 act just as money-maximising. Contrariwise, our
experiment is a one-stage experiment where both parties have complete knowledge of
the decision problem and so it shows that a substantial proportion of people truly have
preferences for opting out.

Our point (iii) is reminiscent of the pretty well established result that females are more giving than males in the Dictator game\cite{EG,AV,DM,HS09,DEJR,D14,KC,B14,BK95,Ca-Ma}. However, it goes beyond it, suggesting that females are not only more sharing than males, but they also have a stronger tendency to exit from a conflictual situation even at a personal cost. While this result is intriguing, we recommend caution on its interpretation. Study 3 suggests that when the cost of the exit option increases, gender differences in taking the exit option tend to disappear.  Further research may help understand how robust is the result that females are more likely than males to exit from a conflict and how far it can be generalised.

We also believe that further research should be devoted to see whether there are behavioral differences between two-person conflicts and N-person conflicts, with $N>2$. If hyper-altruism is partly driven by moral reasoning, as our analysis of free responses suggests, such a difference might exist, since ``harming more people is worse than harming only one person'' and ``helping more people is better than helping only one person''. We tried to handle a similar problem exploring a three-person conflict (Study 4), but there, harming one person was balanced by helping the third person and thus subjects did not really have the opportunity to harm both people at the same time. A posteriori, it is not a big surprise that our experiment was not successful and that behaviour in our three-person conflict turned out to be statistically equivalent to behaviour in the two-person conflicts reported in Studies 1 and 2.   

Finally, we believe that it would be important to understand what psychological consequences such hyper-altruism can have within a person. If a person is available to pay 1 cent to increase the payoff of an anonymous stranger of 1 cent, it is likely that the same person would sacrifice much more to help a closely related person. Such people may thus experience extreme forms of active sacrifice\cite{VanLange97} in their everyday life, such as \emph{unmitigated communion}, that is the extreme focus on others without the balance of a focus on self\cite{Helgeson}. Since unmitigated communion is known to cause anxiety, depressive symptoms, lower self-esteem, and poorer physical health\cite{Drigotas,Fritz98,Helgeson2}, it would be important to understand the extent to which it can be captured by simple economic games such as the ones we have introduced.

\section*{Acknowledgments}
This paper was presented at the Human Cooperation Lab Meeting at the Department of Psychology at Yale University. We thank all participants, in particular Antonio Alonso Ar\'echar, Jillian Jordan, Gordon Kraft-Todd, and David Rand, for numerous useful comments. We thank Giorgia Cococcioni for assistance with coding participant free-responses. The author was partly funded by the Dutch Research Organization (NWO) grant 612.001.352.




\pagebreak
\begin{center}
\begin{huge}
\textbf{Supplementary Information}
\wl
\end{huge}
\end{center}

This Supplementary Information contains two sections. In the first section we provide all the details about the analysis of the free responses of the subjects who participated in our Study 3. In the second section we report the instructions used in our experiments.

\section*{Analysis of free responses}

In order to support our conclusion that altruistic behavior in Study 3 was driven by hyper-altruistic subjects, that is subjects who evaluate other's payoff \emph{strictly} more than their own, rather than by indifferent subjects, who evaluate the other's payoff the same as their own, we asked a  research assistant to code each response from the altruistic participants in Study 3. The coder was not informed about the purpose of the study and the hypothesis and predictions being tested. For each statement, she was asked which of the following five categories best described it:

\begin{enumerate}
\item[] The participant explicitly said that they took the action because that was the right thing to do (Rightness).
\item[] The participant explicitly said that they took the action because the other action was wrong (Wrongness).
\item[] The participant explicitly said that they took the action because they are generous (Generosity).
\item[] The participant explicitly said that they took an action at random, because they were indifferent between the two actions (Indifference).
\item[] The participant said something that is not classifiable in any of the previous categories (Not classifiable).
\end{enumerate}

Below we report all 30 responses. Next to each response, in parenthesis, we report the category to which the response was assigned.

\begin{enumerate}
\item I'd feel bad if I took it (wrongness).
\item it's the holiday season..it is about giving, so I gave. I hope they appreciate it (generosity).
\item i would not feel right taking money from a person (wrongness).
\item I'm a giver, not a taker (generosity).
\item Did not want to be greedy. If there was a 3rd option of keeping my 10c and the other keeping his I would have selected that (indifference).
\item I like giving instead of receiving (generosity).
\item Rather have someone else gain a bonus (generosity).
\item Maybe the other person needs the money more than me and I won't take it from someone else's hand (rightness).
\item Just felt generous (generosity)
\item i just felt like being nice for once (rightness)
\item I wasn't going to take the other person's bonus (wrongness).
\item The other person most likely needs the money more than I do (rightness).
\item Although I considered taking the money, I decided that since it was such a small amount and that I would feel guilty for taking the money, I decided to give up my money (wrongness).
\item It's wrong to take something away from someone else for what amounts to an insignificant gain for myself (wrongness).
\item I would not feel good about myself by taking everything while another person had nothing (wrongness).
\item I thought it would make someone happy and they might need it more than I do (generosity).
\item I want to be good (not classifiable)
\item Better to give than recieve (rightness).
\item I am generous (generosity).
\item I didn't feel right taking all of the money (wrongness).
\item I'd rather give something than take something (generosity)
\item I would much rather give than take. It just doesn't feel right to take something away (wrongness)
\item I feel better not taking away the others money. The benefit is less than the cost of being mean (rightness).
\item I didn't think it was fair to take the other person's money (wrongness).
\item I felt that I wanted to give the amount of the bonus, because I felt it unfair to take all and assume that the other participant would gain more. I generally like to be fair, and hope that both participants receive an equal or substantial amount for the work done. I don't like to be overtly greedy (rightness).
\item Because I didn't want to take from the other participant, and since the only other option was to give my ten cents to him/her, I decided that worked better for my conscience (rightness).
\item It was the easiest choice and no conflict (not classifiable).
\item It was the most kind thing (rightness).
\item I had to make a choice (indifference).
\item I would have felt guilty leaving someone else with nothing but by giving up my 10 cents I feel as if I've done something small but good (wrongness).
\end{enumerate}

\section*{Experimental instructions}

The first screen, where the subjects were asked for their TurkID, was the same for all four studies. After this screen, participants entered the real game, where, in a single screen, we presented the problem and asked to make a decision. After making their decision, subjects entered a standard demographic questionnaire where we asked for their gender, age, education level, and reason for their choice (only in Study 1 and Study 3). Below we report the instructions used in the decision screen for each of the four studies. For each study, we report only the instructions used in the no-exit condition. Those for the other conditions were very similar, a part from the obvious changes.

\emph{Study 1}

You have been paired with another anonymous participant. You both own $\$0.30$ for participating in this HIT. Please choose one of the following alternatives:
\begin{itemize}
\item Donate your $\$0.30$ to the other participant. In this case you end the game with nothing and the other participant ends the game with $\$0.60$.
\item Steal the participation fee from the other participant. In this case you end the game with $\$0.60$ and the other participant with nothing.
\end{itemize}
\wl
\emph{Study 2}

You have been paired with another anonymous participant. You are both given additional $\$0.30$ as a bonus. Please choose one of the following alternatives:
\begin{itemize}
\item Give your $\$0.30$ to the other participant. In this case you end the game with nothing and the other participant ends the game with $\$0.60$.
\item Take the other participant's bonus. In this case you end the game with $\$0.60$ and the other participant with nothing.
\end{itemize}
\wl
\emph{Study 3}

You have been paired with another anonymous participant. You are both given additional $\$0.10$ as a bonus. You can either give your $\$0.10$ to the other participant or take his or her $\$0.10$. In this latter case, the money will be doubled and earned by you. What is your choice?
\begin{itemize}
\item Give your $\$0.10$ to the other participant. In this case you end the game with nothing and the other participant with $\$0.20$.
\item Take $\$0.10$ from the other participant. In this case you end the game with $\$0.30$ and the other participant with nothing.
\end{itemize}
\wl
\emph{Study 4}

You have been grouped together with other two participants, Person A and Person B. You are all given additional $\$0.30$ as a bonus. Please choose one of the following alternatives:
\begin{itemize}
\item Take the $\$0.30$ from Person A and share them with person B. In this case, Person A will finish this task with $\$0$ and you and Person B will finish this task with $\$0.45$.
\item Take the $\$0.30$ from Person B and share them with person A. In this case, Person B will finish this task with $\$0$ and you and Person A will finish this task with $\$0.45$.
\item Give your $\$0.30$ to Person A and Person B. In this case, you will finish this task with $\$0$ and Person A and Person B will finish this task with $\$0.45$.
\end{itemize}

\end{document}